\documentclass[aps,pre,onecolumn,showpacs,superscriptaddress]{revtex4}
\usepackage{epsfig,color,graphics}
\usepackage{amsmath, mathrsfs}
\usepackage{amssymb}
\usepackage{siunitx}
\begin{document}

\title{Synchronous whirling of spinning homogeneous elastic cylinders:
\\ linear and weakly non-linear analyses}

\author{Serge Mora}
\affiliation{Laboratoire de M\'ecanique et G\'enie Civil, Universit\'e de Montpellier and CNRS, France}
\email[Corresponding author: ]{serge.mora@umontpellier.fr}

\date{\today}
\begin{abstract}
  Stationary whirling of slender and homogeneous (continuous) elastic shafts rotating around their axis, with pin-pin boundary condition at the ends, is revisited by considering the complete deformations in the cross section of the shaft. The stability against a synchronous sinusoidal disturbance of any wave length is investigated and the analytic expression of the buckling amplitude is derived in the weakly non-linear regime by considering both geometric and material (hyper-elastic) non-linearities. The bifurcation is super-critical in the long wave length domain for any elastic constitutive law, and sub-critical in the short wave length limit for a limited range of non-linear material parameters.
\end{abstract}
\pacs{46.32.+x,46.25.-y,83.10.Gr, 05.45.-a}

\maketitle
\section{Introduction}

A homogeneous and balanced elastic cylinder rotating around its axis is unstable beyond a critical angular velocity, leading to transverse deformations and whirling if the ends of the cylinder are constraint for instance with bearings. This instability results from the competition between the destabilizing effect of the centrifugal force that tends to drive the cylinder away from the axis of rotation, and the elastic forces opposed to the deformation.

The whirling of rotating cylinders, as well as the propagation of vibrations in the neighborhood of the critical angular velocity, have been extensively investigated in the context of rotor-dynamics \cite{Kramer1993,Genta2005} because of their damaging effects on the smooth running of rotating machinery such  as compressors,  pumps, turbines,  turbochargers, jet engines \cite{Chen2005}. Understanding the stability  of spinning shafts and their post-buckling behavior is crucial for the success in the design of this kind of rotating systems.

While most of the studies have dealt with small deformations linearized at leading order \cite{Ehrich1964}, few studies have considered non-linear effects \cite{Noah1995,Yamamoto2012}. The non-linear dynamic behaviour of a uniform, slender rotating shaft made of a viscoelastic material with external damping mechanism has been studied  by considering geometric non-linearities resulting from large transverse displacements \cite{Shaw1989,Kurnik1994,Hosseini2013}. Using the center manifold technique \cite{Henry1981} and the normal form method, the effects of external and internal damping on the whirling of rotating shafts have been investigated in terms of Hopf or double eigenvalues bifurcations. By pushing expansions up to order 2 in terms of the characteristic magnitude of the infinitesimal strain, $\varepsilon$, but Hookean elasticity for the strain-stress relation, the whirling amplitude in steady state configurations have been computed as the radius of a limit cycle in phase portraits \cite{Hosseini2013}. However, the intrinsic non-linear features of material constitutive law have been neglected in these studies. Indeed, order $\varepsilon^2$ in the expansion of the governing equations originates both from geometrical non-linearities (arising from the expression of the local curvature of the center line of the cylinder) together with non-linearities in the constitutive law of the elastic material. These last non-linearities are essential in order to fulfilled the requirement of material objectivity \cite{Ogden1984}. 

An expansion of the bending energy based on a scalar non-linear constitutive law \cite{Haslach1985} has been proposed in order to calculate non-synchronous whirling of rotating shafts \cite{Cveticanin1998}. Because of the scalar features of the constitutive law used by the author, this approach is limited to deformation with large wave length (compared with the radius of the shaft) and the issues related to Poisson effect are ignored. In addition, the rotating shaft was supposed to be not extensible which is not relevant for pin-pin ends since the extension of the center-line with  pin-pin ends is of order $\varepsilon^2$ and cannot be neglected.

A linear analysis of the whirling bifurcation of infinite rotating cylinders under axial tension has been developed in \cite{Ogden1980a}, based on non-linear constitutive equations in three dimensions so that this analysis is relevant for any wave length of the deformation, but the non-linear analysis is still missing.  In previous papers \cite{Richard2018,Richard2019}, the bifurcations of spinning undeformable shafts, surrounded by a compliant elastic layer, have been investigated both in the linear and the non-linear regimes, under plane strain assumption. \\

In this paper, a non-linear analysis of the {\it stationary} whirling of {\it homogeneous} rotating cylinders is developed, based on the hypothesis of negligible external damping \cite{Ehrich1964} so that the system is conservative.  The steady states are reached once transient vibrations are damped thanks to dissipative processes (internal damping) occurring inside the elastic material. The cylinders are supposed to be slender, their length $L$ being far larger than the radius $r_0$. The elastic material is assumed to be isotropic and incompressible. The buckling amplitude of synchronous and steady sinusoidal perturbations of any wave length is calculated without any further assumption for the constitutive law of the elastic material. The analysis relies on the complete three dimensional equations so that the results are relevant for any wave length of the whirling, including wave length of the same order of magnitude as the radius of the shaft.
The complete (non-linear) equations governing the equilibrium steady states are derived in Section \ref{sec : base equations}. A Lagrange multiplier accounts for the incompressibility constraint and the equations for the three components of displacement field are established in strong form. Section \ref{sec : linear} is devoted to the linear stability analysis. The critical angular velocity is found to depend on the shear modulus of the elastic material, its mass density, the radius of the rotating cylinder, and in a non trivial manner on the ratio of the wave length of the deformation to the radius of the cylinder. 
The weakly non-linear analysis of the bifurcation is carried out in Section \ref{sec : non linear}. The bifurcation is found to be super-critical for neo-Hookean materials, and can  be sub-critical at small wave length for particular constitutive laws. Predictions of sections \ref{sec : linear}-\ref{sec : non linear} are checked in Section \ref{sec : FEM} by means of numerical simulations based on the Finite Element Method. The last part (Section \ref{sec : conclusion}) of the paper is devoted to a conclusion.

\section{Equilibrium equations based on a finite strain theory} \label{sec : base equations}

In this section the non-linear equations governing the equilibrium (steady) configurations of a rotating elastic cylinder are derived, considering an arbitrary
hyper-elastic incompressible isotropic material.

\begin{figure}[!h]
\begin{center}
\includegraphics[width=0.55\textwidth]{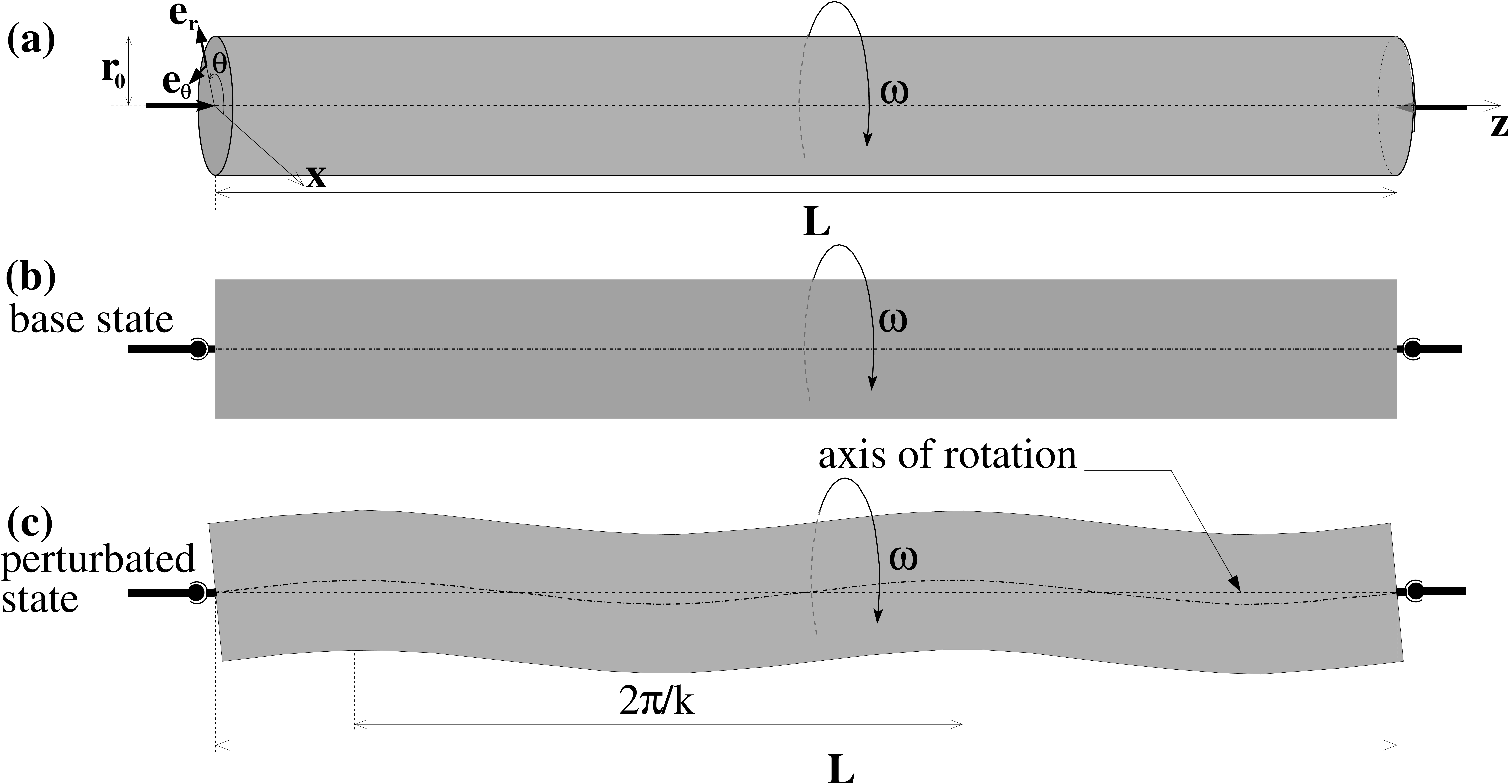}
\end{center}
\caption{Sketches of an elastic cylinder of length $L$ and radius $r_0$ rotating around its axis with the angular velocity $\omega$. The two ends of the cylinder are pinned at the axis. (a) Three dimensional view of the reference (unbuckled) configuration. (b) Side view of this reference configuration. (c) Side view of a perturbation of characteristic wave vector $k$ parallel to the axis.} \label{fig : scheme}
\end{figure}

Let $r_{0}$ denote the radius of the undeformed cylinder, $\rho$ its
mass density and $\mu$ its initial shear modulus, {\it i.e.} the
shear modulus for infinitesimal strain.  The cylinder is spun with an
angular velocity $\omega$ about its axis, as sketched in
Figure \ref{fig : scheme}.  
In the co-rotating frame, both the elastic force and the centrifugal
force are conservative.  The equilibrium can therefore be derived from
the condition that the total potential energy is stationary.  The
position $\mathbf{R}$ of a material point in the deformed configuration is
given as a map ${\mathbf{R}}(\mathbf{r})$ in terms of the position
$\mathbf{r}$ in the undeformed configuration.  For an isotropic and
incompressible elastic material, the strain energy density is a
function of the two first invariants, $I_1$ and $I_2$, of Green's
deformation tensor $\mathbf{C}=\mathbf{F}^T\cdot\mathbf{F}$, where
$\mathbf{F}=\partial \mathbf{R}/\partial \mathbf{r}$ is the
deformation gradient:
\begin{equation}
  \begin{array}{lll}
    I_1 & = & \mathrm{tr}~ \mathbf{C} - 3,\\
    I_2 & = & \frac{1}{2}\left(\left(\mathrm{tr}~ \mathbf{C}\right)^2 -
    \mathrm{tr}~\left(\mathbf{C}^2\right) \right) - 3.
  \end{array} \label{eq:invariants}
\end{equation}
The strain energy density is then written as $\mu\, W(I_1,I_2)$ where
$W$ is the dimensionless strain energy density.    For the
strain energy $\mu\, W(I_1,I_2)$ to be consistent with the initial
shear modulus $\mu$, the following normalization condition must be
enforced:
\begin{equation}
  \frac{\partial W}{\partial I_1}(0,0) + \frac{\partial W}{\partial I_2}(0,0)=\frac{1}{2}.
  \label{eqn : tangent modulus}
\end{equation}
For an incompressible neo-Hookean
solid~\citep{Ogden1984,Macosko94} and for an incompressible Mooney-Rivlin
solid \citep{Mooney1940,Rivlin1948}, the dimensionless strain energy density are respectively
$W=\frac{1}{2}(I_1-3)$ and 
$W=\frac{1}{2}\left(\beta(I_1-3)+(1-\beta)(I_2-3)\right)$, with $\beta$ a
material constant in the range $[0;1]$.

Incompressibility of the elastic material imposes the condition
$\mathcal{D}(\mathbf{r}) = 1$, where $\mathcal{D}=\det \mathbf{F}$ is
the Jacobian of the transformation.
To characterize equilibrium configurations, we seek stationary points of the augmented energy
\begin{equation}
  {\cal E}=\int_{ 0<r<r_0;~0<z<L} \mathrm{d}\mathbf{r}
  \,\left( \mu\,{W}(I_1,I_2) -\frac{1}{2}\rho\,\omega^2\,
  \left(\mathbf{R} \cdot \mathbf{R}-(\mathbf{R}\cdot \mathbf{e}_z)^2\right)
  +\mu\,q\,({\cal D}-1) \right).
  \label{eqn : energy general}
\end{equation}
The terms in the integrand are the strain energy, the potential of the
centrifugal force, and the Lagrange term taking care of the
incompressibility constraint ${\cal D}= 1$ by means of a Lagrange multiplier
$q(\mathbf{r})$. From Eq.~\ref{eqn : energy general}, the equilibrium of the system is
governed by the two dimensionless parameters in the problem, namely  $\alpha=\rho \, r_0^2\, \omega^2/ \mu$ and the ratio $L/r_0$.
We use cylindrical coordinates, with $r$ the distance to the axis,  $\theta$ the angle and $z$ the height in the unperturbed state (Figure \ref{fig : scheme}).  Let $\mathbf{e_r}$, $\mathbf{e_\theta}$,$\mathbf{e_z}$ be the orthonormal basis vectors associated with coordinates $r$, $\theta$ and $z$ respectively. In
the deformed configuration, the
position $\mathbf{R}$ of a material point is $\mathbf{R}=\mathbf{r}+u(r,\theta,z)\mathbf{e_r}+v(r,\theta,z)\mathbf{e_\theta}+w(r,\theta,z)\mathbf{e_z}$ and the  deformation gradient $\mathbf{F} =\nabla\mathbf{R} ({\mathbf{r}})$ is:
\begin{equation}
  \mathbf{F}=\left(
  \begin{array}{ccc}
    1+u_{,r}&(u_{,\theta}-v)/r&u_{,z}\\ \\
    v_{,r}&(v_{,\theta}+u)/r+1&v_{,z}\\ \\
    w_{,r}&w_{,\theta}/r&w_{,z}+1
    \end{array}
  \right),
  \label{eqn : 3d F}
\end{equation}
where a comma in subscript denotes a partial derivative.
Expressions of ${\cal D}$, $I_1$ and $I_2$ are directly deduced from Eq.~\ref{eqn : 3d F}.

The equilibrium equations are derived from the condition that the
first variation of Eq.~\ref{eqn : energy general} with respect to the
unknowns $u(r,\theta,z)$, $v(r,\theta,z)$, $w(r,\theta,z)$ and $q(r,\theta,z)$ is zero.
Let $\mathbf{t}=(u,v,w,q)$ denote the collection of unknowns, and
$\delta \mathbf{t}=(\delta u,\delta v, \delta w, \delta q)$ a virtual
displacement that is kinematically admissible (abbreviated as `k.a.'),
 as imposed by the boundary conditions.  The field $\mathbf{t}(r,\theta,z)$ is a solution of the problem
if
\begin{equation}
  \forall \delta \mathbf{t}\;\mathrm{k.a.}, \quad  D{\cal E}(\alpha,\mathbf{t})\left[\delta \mathbf{t} \right]=0.
  \label{eqn : general1}
\end{equation}
$D{\cal E}(\alpha,\mathbf{t})\left[\delta \mathbf{t} \right]$
denotes the first variation of the energy evaluated in the
configuration $\mathbf{t}$ with an increment $\delta \mathbf{t}$, also
known as the first G\^ateaux derivative of the functional ${\cal E}$
\citep{gateaux}. Note that the dependence of ${\cal E}$ with $L/r_0$
is not explicitly written in Eq.~\ref{eqn : general1} because it is a fixed parameter in the system, contrary to $\alpha$.

Defining
\begin{equation}
  {\cal G}=r\left\{W(I_1,I_2)+q({\cal D}-1) -\frac{1}{2}\frac{\alpha}{r_0^2}\left((r+u)^2+v^2\right)\right\}
\end{equation}
and integrating by parts Eq.~\ref{eqn : general1}, we obtain the equations in the interior of the body as
\begin{eqnarray}
  {\cal D}-1&=&0, \label{eqn : 3d CP0 brute}\\
\frac{\partial {\cal G}}{\partial u}-\frac{\partial}{\partial r}\left(\frac{\partial {\cal G}}{\partial u_{,r}} \right)-\frac{\partial}{\partial \theta}\left(\frac{\partial{\cal G}}{\partial u_{,\theta}} \right)-\frac{\partial}{\partial z}\left(\frac{\partial{\cal G}}{\partial u_{,z}} \right)&=&0, \label{eqn : 3d CP1 brute}\\
\frac{\partial {\cal G}}{\partial v}-\frac{\partial}{\partial r}\left(\frac{\partial{\cal G}}{\partial v_{,r}} \right)-\frac{\partial}{\partial \theta}\left(\frac{\partial{\cal G}}{\partial v_{,\theta}} \right)-\frac{\partial}{\partial z}\left(\frac{\partial{\cal G}}{\partial v_{,z}} \right)&=&0, \label{eqn : 3d CP2 brute}\\
\frac{\partial {\cal G}}{\partial w}-\frac{\partial}{\partial r}\left(\frac{\partial{\cal G}}{\partial w_{,r}} \right)-\frac{\partial}{\partial \theta}\left(\frac{\partial{\cal G}}{\partial w_{,\theta}} \right)-\frac{\partial}{\partial z}\left(\frac{\partial{\cal G}}{\partial w_{,z}} \right)&=&0. \label{eqn : 3d CP3 brute}
\end{eqnarray}
The first equation (Eq.~\ref{eqn : 3d CP0 brute}) is the incompressibility constraint and the
three other equations (Eqs.~\ref{eqn : 3d CP1 brute}-\ref{eqn : 3d CP3 brute}) are the equilibrium in the radial,
circumferential and longitudinal directions, respectively.  These equations are
complemented by the condition of zero traction at the lateral boundary
$r=r_0$,
\begin{equation}
 \left. \frac{\partial{\cal G}}{\partial u_{,r}}\right|_{r=r_0}=\left. \frac{\partial{\cal G}}{\partial v_{,r}}\right|_{r=r_0}=\left. \frac{\partial{\cal G}}{\partial w_{,r}}\right|_{r=r_0}=0.
   \label{eqn : 3d BC3 brute}
\end{equation}
In addition, the pin-pin condition at the ends imposes:
\begin{equation}
   u(0,\theta,z)=v(0,\theta,z)=w(0,\theta,z)=0 \mbox{ for } z=0 \mbox{ and } z=L
\label{eqn : boundary ends 1}
\end{equation}
and
\begin{equation}
 \frac{\partial{\cal G}}{\partial u_{,z}}=\frac{\partial{\cal G}}{\partial v_{,z}}=\frac{\partial{\cal G}}{\partial w_{,z}}=0 \mbox{ for } z=0 \mbox{ and } z=L.
   \label{eqn : boundary ends 2}
\end{equation}
The three last boundary conditions (Eqs.~\ref{eqn : boundary ends 2}) originate from the variation of the augmented energy at the vicinity of the ends.
Since end effects are expected to spread in a domain of characteristic size $r_0$, their relative contribution to the total augmented energy is of order $r_0/L$. Hence, within the hypothesis of a slender shaft ($r_0 \gg L$), boundary conditions Eqs.~\ref{eqn : boundary ends 2} is negligible. This simplification makes possible the harmonic decomposition of the deformation (with unique wave length and unique circumferential wave number, see Section \ref{sec : linearization}).\\

The equilibrium configurations are the solutions of the system formed by Eqs.~\ref{eqn : 3d CP0 brute}-\ref{eqn : boundary ends 1}. Because of non-linearities in the equations, the analytic resolution is out of reach. In Section \ref{sec : unbuckled} the system is resolved in the reference (undeformed) configuration. Then the magnitude of the displacement is assumed to scale as a small parameter, $\varepsilon$, so that it is infinitely smaller than the other length scales ($r_0$ and $L$). Eqs.~\ref{eqn : 3d CP0 brute}-\ref{eqn : boundary ends 1} is resolved at linear order (order $\varepsilon$) in Section \ref{sec : linearization}, and at order $\varepsilon^2$ in Section \ref{sec : non linear}. Finally, they are solved numerically by means of finite elements in Section \ref{sec : FEM}. 

\section{linear bifurcation analysis} \label{sec : linear}
\subsection{Unbuckled solution} \label{sec : unbuckled}
We start by analyzing the unbuckled configuration (base state), and
label all quantities relevant to it using a subscript `$0$'. In this configuration, $u_{0}(r,\theta,z)=0$, $v_{0}(r,\theta,z)=0$ and $w_{0}(r,\theta,z)=0$.  The Lagrange multiplier $q$ is found from the radial equilibrium Eq.~\ref{eqn : 3d CP1 brute}
and Eq.~\ref{eqn : 3d BC3 brute} as
\begin{equation} 
q_0=\frac{\alpha}{2}\left(1-(r/r_0)^2\right)+\beta-2.
  \label{eqn : base lagrange multiplier}
\end{equation}
Altogether, the unbuckled solution of Eqs.~\ref{eqn : 3d CP0 brute}-\ref{eqn : 3d BC3 brute} is written as $\mathbf{t}_0=(u_0,v_0,,w_0,q_0)$.

\subsection{Linearization of the equations} \label{sec : linearization}
A small perturbation is added to the unbuckled solution, and the
equations of Section~\ref{sec : base equations} are linearized with
respect to the amplitude of the perturbation,
\begin{equation}
\mathbf{t}=\mathbf{t}_0+\varepsilon \mathbf{t}_1
    =\left(\varepsilon u_1(r,\theta,z),\varepsilon v_1(r,\theta,z),\varepsilon w_1(r,\theta,z),q_0(r,\theta,z)+\varepsilon q_1(r,\theta,z)\right).
	\label{eq:order1expansion}
\end{equation}
We first assume a harmonic $\theta$ and $z$ dependence of any variation of the perturbation of $u$, $v$, $w$ and $q$:
\begin{equation}
\left\{ \begin{array}{l}
u_1=u^+_1(r,\theta,z)={\cal R}e \left(f_u(r)e^{\mathrm{i}\theta+\mathrm{i}kz} \right) \\
v_1=v^+_1(r,\theta,z)={\cal R}e \left(-\mathrm{i}f_v(r)e^{\mathrm{i}\theta+\mathrm{i}kz} \right)  \\
w_1=w_1^+(r,\theta,z)={\cal R}e \left(-\mathrm{i}f_w(r)e^{\mathrm{i}\theta+\mathrm{i}kz} \right) \\
q_1=q^+_1(r,\theta,z)={\cal R}e \left(f_q(r)e^{\mathrm{i}\theta+\mathrm{i}kz} \right)
\end{array} \right.
\label{eqn : helicoidal +}
\end{equation}
where $k$ is the axial wave number.
${\cal R}e$ denotes the real part.  The conventional complex
factor $(-\mathrm{i})$ has been included for convenience, anticipating on the fact that the phase of $v_{1}$ and $w_1$ are shifted by $\pi/2$ compared to the
phase of the two other unknowns.
At linear order in $\varepsilon$,
Eqs.~\ref{eqn : 3d CP0 brute}-\ref{eqn : 3d BC3 brute}  yield respectively:
\begin{equation}
r{{d f_u}\over{dr}}+f_u+kr f_w+f_v=0
   \label{eqn : 3d CP0 lin}
\end{equation}
\begin{equation}
-{{d^2 f_u}\over{d r^2}}-\frac{1}{r}{{d f_u}\over{dr}}+(k^2+\frac{2}{r^2})f_u+\left({{\alpha }\over{r_{0}^2}}+\frac{2}{r^2}\right)f_v+{{\alpha krf_w}\over{r_{0}^2}}-{{d f_q}\over{dr}} =0
      \label{eqn : 3d CP1 lin}
\end{equation}
\begin{equation}
\left(\frac{2}{r}+{{\alpha r}\over{r_{0}^2}}\right)f_u-r{{ d^2 f_v}\over{dr^2}}-{{d f_v}\over{dr}}+(\frac{2}{r}+k^2r)f_v +f_q=0
\end{equation}
\begin{equation}
{{\alpha rf_u}\over{r_{0}^2}}-\frac{1}{k}{{d^2 f_w}\over{dr^2}}-\frac{1}{rk}{{d f_w}\over{dr}}+\left(\frac{1}{kr^2}+k\right)f_w+f_q=0.
       \label{eqn : 3d CP3 lin}
\end{equation}
  The boundary conditions Eqs.~\ref{eqn : 3d CP0 brute}-\ref{eqn : 3d CP3 brute} at order $\varepsilon$ are respectively:
  \begin{equation}
r_0{{d f_u}\over{dr}}-kr_0 f_w-f_v-f_u +r_0f_q=0
~~~~~~ \mbox{ at } r=r_0,
    \label{eqn : 3d BC1 lin}
\end{equation}
\begin{equation}
    r_0{{d f_v}\over{dr}}-f_v -f_u=0 ~~~~~~ \mbox{ at } r=r_0,
    \label{eqn : 3d BC2 lin}
\end{equation}
\begin{equation}
    r_0 {{d f_w}\over{dr}}-kr_0 f_u=0 ~~~~~~ \mbox{ at } r=r_0.
    \label{eqn : 3d BC3 lin}
\end{equation}
After the elimination of $f_v$, $f_w$ and $f_q$ in  Eqs.~\ref{eqn : 3d CP0 lin}-\ref{eqn : 3d CP3 lin}, one obtains an order 6 differential equation for $f_u$:
\begin{equation}
  \begin{split}
    r^5 {{d^6 f_u}\over{d r^6}} +&9 r ^4 {{d^5 f_u}\over{d r^5}} f_u+\left( 9 r^3-3 k^2 r^5\right) {{d^4 f_u}\over{d r^4}}+\left(-18 k^2 r^4-12 r^2\right){{d^3 f_u}\over{d r^3}}+\left(3 k^4 r^5-9  k^2 r^3+9 r\right) {{d^2 f_u}\over{d r^2}}\\
    &+\left(9 k^4 r^4+9 k^2 r^2-9\right) {{d f_u}\over{d r}} -k^6 r^5 f_u=0 
  \end{split}
  \label{eqn : 3d equa diff}
  \end{equation}
and, after substitutions in Eqs.~\ref{eqn : 3d BC1 lin}-\ref{eqn : 3d BC3 lin}, one obtains the boundary conditions at $r=r_0$ in term of $f_u$:
\begin{equation}
  \begin{split}
-r_0^3 {{d^4 f_u}\over{d r^4}} f_u-6  r_0^2 {{d^3 f_u}\over{d r^3}}+ \left(2 k^2 r_0^3-3 r_0\right) {{d^2 f_u}\over{d r^2}}+\left(14 k^2 r_0^2+3\right) {{d f_u}\over{d  r}}+\left(-k^4 r_0^3-4 \alpha k^2  r_0\right) f_u=0
    \label{eqn : 3d BC1}
    \end{split}
\end{equation}
\begin{equation}
  \begin{split}
    -r_0^5 {{d^6 f_u}\over{d r^6}}&-9  r_0^4 {{d^5 f_u}\over{d r^5}} + \left(2 k^2 r_0^5-10 r_0^3\right) {{d^4 f_u}\over{d r^4}} +\left(16 k^2 r_0^4+6 r_0^2\right)  {{d^3 f_u}\over{d r^3}}+\left(-k^4  r_0^5+16 k^2 r_0^3-12 r_0\right) {{d^2 f_u}\over{d r^2}}\\
    &+\left(-7 k^4 r_0^4-8 k^2 r_0^2+12 \right) {{d f_u}\over{d r}}-5 k^ 4 r_0^3 f_u=0
  \label{eqn : 3d BC2}
  \end{split}
\end{equation}

\begin{equation}
  \begin{split}
    r_0^5 {{d^6 f_u}\over{d r^6}} &+9 r_0 ^4 {{d^5 f_u}\over{d r^5}} +\left( 10 r_0^3-2 k^2 r_0^5\right) {{d^4 f_u}\over{d r^4}} +\left(-16 k^2 r_0^4-6 r_0^2\right){{d^3 f_u }\over{d r^3}} +\left(k^4 r_0^5-24 k^ 2 r_0^3+12 r_0\right) {{d^2 f_u}\over{d r^2}} \\
    &+\left(7 k^4 r_0^4-12\right) {{d f_u}\over{d r}} -3 k^4 r_0^3 f_u=0.
  \label{eqn : 3d BC3}
  \end{split}
\end{equation}

\subsection{General solution}

\begin{figure}
\begin{center}
\includegraphics[width=0.45\textwidth]{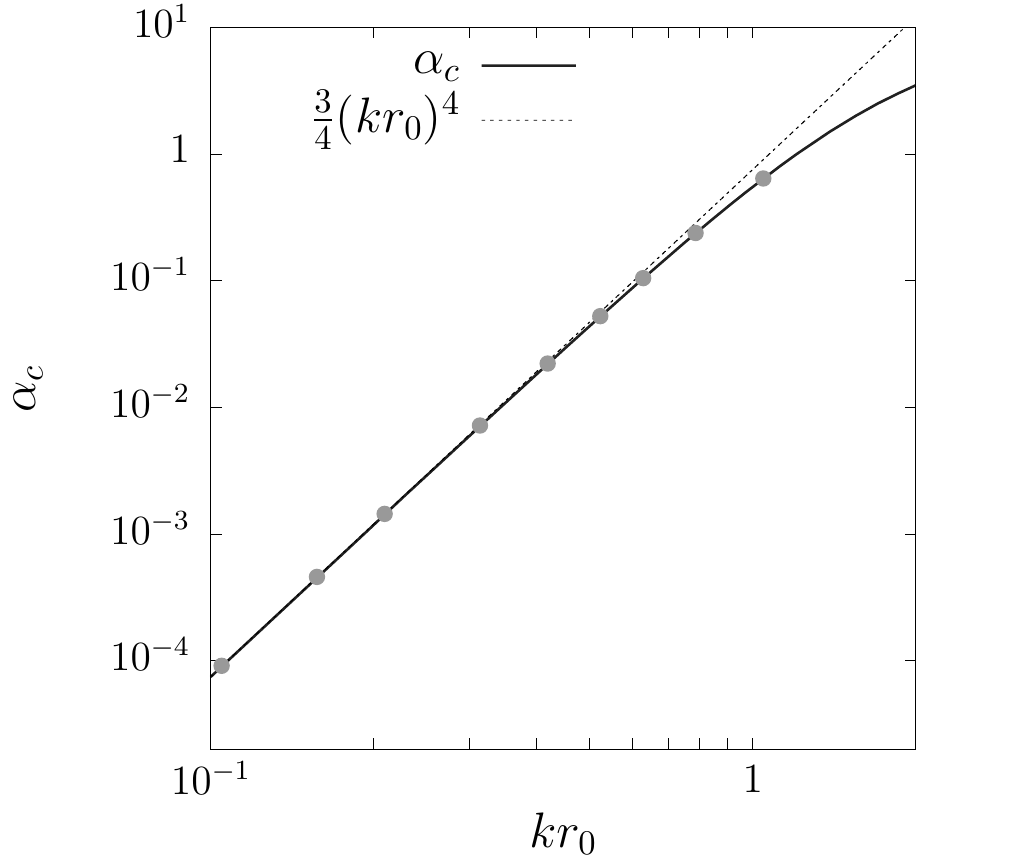}
\end{center}
\caption{Solid line: Critical value of $\alpha$  at the instability onset, as a function of  $k$, calculated from Eq.~\ref{eqn : alphac} at order 30 in $kr_0$ (higher orders in the expansion lead to indistinguishable curves). Dashed line: First term in the expansion of $\alpha_c$ with respect to $kr_0$. Filled circles:  Threshold $\alpha^*$ obtained from FEM simulations (see Eq.~\ref{eqn : star} of Section \ref{sec : FEM}).} \label{fig : linear threshold}
\end{figure}

Let $s_1(kr)$,  $s_2(kr)$ and $s_3(kr)$ be three independent solutions of Eqs.~\ref{eqn : 3d equa diff}-\ref{eqn : 3d BC3} that do not diverge, as well as their first derivative, at $r=0$. These solutions are sought as series expansions in the form:  
\begin{equation}
  s_i(kr)=\sum_{m=0}^\infty a_m(kr)^m.
\end{equation}
The condition for $s_i(kr)$ to be a solution of Eq.~\ref{eqn : 3d equa diff} is, for $m\ge 6$:
\begin{equation}
 a_{m-6}-3a_{m-4}(m-4)(m-2)+3a_{m-2}m(m-2)^2(m-4) -a_m(m-4)(m-2)^2m^2(2+m)=0,
  \end{equation}
where $a_0$, $a_2$ and $a_4$ are constants that are not fixed up to now. Coefficients $a_m$ with an odd index have to be 0. In order to build three independent solutions of Eq.~\ref{eqn : 3d equa diff}, we choose $a_0=1$, $a_2=a_4=0$ for $s_1(kr)$ ;
$a_0=a_4=0$ and $a_2=1$ for $s_2(kr)$ ; and
$a_0=a_2=0$ and $a_4=1$ for $s_3(kr)$. Writing now the general solution $f_u(r)$ of Eq.~\ref{eqn : 3d equa diff} as:
\begin{equation}
  f_u(r)=As_1(kr)+Bs_2(kr)+Cs_3(kr),
\end{equation}
and substituting this expression in the boundary conditions Eqs.~\ref{eqn : 3d BC1}-\ref{eqn : 3d BC3}, one gets a linear system of 3 homogeneous equations with three unknowns $A$, $B$ and $C$.  The condition for a non-zero deformation, {\it i.e.} $(A,B,C)\ne(0,0,0)$, is obtained by imposing the determinant of the linear system to be zero, leading to the condition for $\alpha$,  $\alpha=\alpha_c$ with $\alpha_c$:
\begin{equation}
  \alpha_c=\frac{3}{4}(kr_0)^4-\frac{5}{24}(kr_0)^6 -\frac{19}{1536}(kr_0)^8 + \cdots
  \label{eqn : alphac}
\end{equation}
Higher orders in the expansion can be calculated as well. For $\alpha=\alpha_c$, the system is neutrally stable against a perturbation of wave number $k$. $\alpha_c$ is plotted as a function of $kr_0$ in Figure \ref{fig : linear threshold}.  
Taking the first term in the expansion Eq.~\ref{eqn : alphac}, one recovers the well known expression of the linear threshold calculated in the long wave length limit  in the framework of Hookean elasticity, $\alpha_c=\frac{3}{4}(kr_0)^4$ (see the dashed line in Figure \ref{fig : linear threshold}). 

The expressions of functions $f_u(r)$, $f_v(r)$, $f_w(r)$ and $f_q(r)$, with the condition $f_u(r_0)=\xi$ ($\varepsilon \xi$ will be referred as the buckling amplitude)  are: 
\begin{eqnarray}
  \frac{f_u(r)}{\xi}&=&1+\left(1-\frac{r^2}{r_0^2}\right)\frac{(kr_0)^2}{4}-\left(3-\frac{4r^2}{r_0^2}+\frac{r^4}{r_0^4}\right)\frac{3(kr_0)^4}{64}+\left(23-\frac{39r^2}{r_0^2}+\frac{21r^4}{r_0^4}-\frac{5r^6}{r_0^6}\right)\frac{(kr_0)^6}{2304}+\cdots\\
  \frac{f_v(r)}{\xi}&=&-1-\left(1+\frac{r^2}{r_0^2}\right)\frac{(kr_0)^2}{4}+\left(9+\frac{12r^2}{r_0^2}-\frac{r^4}{r_0^4}\right)\frac{(kr_0)^4}{64}-\left(23+\frac{135r^2}{r_0^2}-\frac{39r^4}{r_0^4}+\frac{r^6}{r_0^6}\right)\frac{(kr_0)^6}{2304}+\cdots\\
  \frac{f_w(r)}{\xi}&=&\frac{r}{r_0}(kr_0)-\left(3\frac{r}{r_0}-\frac{r^3}{r_0^3}\right)\frac{(kr_0)^3}{4}+\left(\frac{7r}{r_0}-\frac{4r^3}{r_0^3}+\frac{r^5}{r_0^5}\right)\frac{(kr_0)^5}{64}+\cdots\\
  f_q(r)&=&\frac{\xi}{r_0}\left\{ \frac{r}{r_0}(kr_0)^2-\left(4\frac{r}{r_0}-\frac{r^3}{r_0^3}\right)\frac{(kr_0)^4}{8}-\left(\frac{39r}{r_0}-\frac{42r^3}{r_0^3}-\frac{r^5}{r^5_0}\right)\frac{(kr_0)^6}{192}+\cdots \right\}
\end{eqnarray}

Indeed, the form of Eq.~\ref{eqn : helicoidal +} corresponds to a right helical deformation along the z axis. The left helical deformation can be deduced from the previous one (with the transformation $k \leftrightarrow -k$) :

\begin{equation}
\left\{ \begin{array}{l}
u_1=u^-_1(r,\theta,z)={\cal R}e \left(f_u(r)e^{\mathrm{i}\theta-\mathrm{i}kz} \right) \\
v_1=v^-_1(r,\theta,z)={\cal R}e \left(-\mathrm{i}f_v(r)e^{\mathrm{i}\theta-\mathrm{i}kz} \right)  \\
w_1=w_1^-(r,\theta,z)={\cal R}e \left(\mathrm{i}f_w(r)e^{\mathrm{i}\theta-\mathrm{i}kz} \right) \\
q_1=q^-_1(r,\theta,z)={\cal R}e \left(f_q(r)e^{\mathrm{i}\theta-\mathrm{i}kz} \right)
\end{array} \right.
\label{eqn : helicoidal -}
\end{equation}

Up to now, the boundary conditions at the ends of the cylinder, Eq.~\ref{eqn : boundary ends 1}, have not been taken into account. In that case, the general solution at linear order of the problem consists of any linear combinations of the solutions Eqs.~\ref{eqn : helicoidal +} and \ref{eqn : helicoidal -}. Imposing now the boundary condition Eq.~\ref{eqn : boundary ends 1} yields the unique (up to the buckling amplitude) solution of the complete problem at linear order:
\begin{eqnarray}
  u_1&=&f_u(r)\sin \theta \sin kz \nonumber\\
  v_1&=&-f_v(r) \cos \theta \sin kz \nonumber\\
  w_1&=&-f_w(r) \sin \theta \cos kz \label{eqn : t linear}\\
  q_1&=&f_q(r) \sin \theta \sin kz \nonumber
\end{eqnarray}
with $k=n\pi/L$ and $n$ an integer. Hence, at linear order, only discrete values of wave number $k$ are admissible for the system to be neutral against sinusoidal perturbations, which corresponds to discrete values of the control parameter $\alpha$.

In the following, we start from a value of $\alpha$ at which the system is neutrally stable ($\alpha=\alpha_c$), and we consider a quasi-static increase of $\alpha$. The deformation is not harmonic anymore, and we calculated the expression of the corresponding mode, including the buckling amplitude.

\section{Weakly non-linear analysis} \label{sec : non linear}

\subsection{Introduction}
In this section, we carry out a Koiter expansion \citep{Koiter-On-the-stability-of-an-elastic-equilibrium-1945,Hutchinson-Imperfection-sensitivity-of-externally-1967,Hutchinson-Koiter-Postbuckling-theory-1970,Budiansky-Theory-of-buckling-and-post-buckling-1974,Peek-Triantafyllidis-Worst-shapes-of-imperfections-1992,Peek-Kheyrkhahan-Postbuckling-behavior-and-imperfection-1993,Heijden2009koiter} of the bifurcated solution in the vicinity of a bifurcation point. The displacement field and the Lagrange multiplier are
expanded to order 3 in terms of an arc-length parameter $\varepsilon$ defined as  \citep{Chakrabarti2018}:
\begin{eqnarray}
        \alpha & =&\alpha_{\mathrm{c}} +\alpha_2 \varepsilon^2 \label{eqn : definition of alpha2}\\
        \mathbf{t}(\alpha) & = &\mathbf{t}_{0}(\alpha) + \varepsilon\mathbf{t}_{1} + \varepsilon^2\mathbf{t}_{2} +\varepsilon^3\mathbf{t}_{3}+ \cdots \label{eqn : definition of xi}
\end{eqnarray}
where $\alpha_c$ is the critical dimensionless angular
velocity determined from the linear bifurcation analysis, see Eq.~\ref{eqn : alphac}.
The base solution $\mathbf{t}_0$ depends on the load $\alpha$ through $q_0$.  The first-order correction $\mathbf{t}_1$ is the linear mode calculated in Section~\ref{sec : linear} and normalized so that the buckling amplitude is $\varepsilon \xi$.

\subsection{Second-order correction to the displacement} \label{sec : second order}

The second order displacements $\mathbf{t}_2=(u_{2},t_{2},z_2,q_{2})$
results from the non-linear interaction of the linear mode
$\mathbf{t}_{1}$ with itself.  As a result, it involves a superposition of Fourier modes having wave numbers $\pm k$  with respect to the variable $z$, and  circumferential wave numbers $\pm 1$.
Hence,  we seek the second-order correction $\mathbf{t}_{2}$ to the displacement as:
\begin{equation}
  \begin{split}
    u_2(r,\theta,z)=&g_{u1}(r)+g_{u2}(r)\sin(2kz)+g_{u3}(r)\cos(2kz)+g_{u4}(r)\sin(2\theta)+g_{u5}(r)\cos(2\theta)+g_{u6}\sin(2\theta)\sin(2kz)\\& +g_{u7}\sin(2\theta)\cos(2kz)+g_{u8}(r)\cos(2\theta)\sin(2kz)+g_{u9}(r)\cos(2\theta)\cos(2kz)  \label{eqn : gu}
    \end{split}
\end{equation}

\begin{equation}
  \begin{split}
    v_2(r,\theta,z)=&g_{v1}(r)+g_{v2}(r)\sin(2kz)+g_{v3}(r)\cos(2kz)+g_{v4}(r)\cos(2\theta)+g_{v5}(r)\sin(2\theta)+g_{v6}(r)\cos(2\theta)\sin(2kz)\\&+g_{v7}(r)\cos(2\theta)\cos(2kz)+g_{v8}\sin(2\theta)\sin(2kz)+g_{v9}\sin(2\theta)\cos(2kz)  \label{eqn : gv}
    \end{split}
\end{equation}

\begin{equation}
  \begin{split}
    w_2(r,\theta,z)=&g_{w1}(r)+g_{w2}(r)\cos(2kz)+g_{w3}(r)\sin(2kz)+g_{w4}(r)\sin(2\theta)+g_{w5}(r)\cos(2\theta)+g_{w6}\sin(2\theta)\cos(2kz)\\&+g_{w7}\sin(2\theta)\sin(2kz)+g_{w8}(r)\cos(2\theta)\cos(2kz) +g_{w9}(r)\cos(2\theta)\sin(2kz)  \label{eqn : gw}
    \end{split}
\end{equation}

\begin{equation}
  \begin{split}
    q_2(r,\theta,z)=&g_{q1}(r)+g_{q2}(r)\sin(2kz)+g_{q3}(r)\cos(2kz)+g_{q4}(r)\sin(2\theta)+g_{q5}(r)\cos(2\theta)+g_{q6}\sin(2\theta)\sin(2kz)\\& +g_{q7}\sin(2\theta)\cos(2kz)+g_{q8}(r)\cos(2\theta)\sin(2kz)+g_{q9}(r)\cos(2\theta)\cos(2kz).  \label{eqn : gq}
    \end{split}
\end{equation}

The calculation of the deformation at order $\varepsilon^2$ requires to take into account a series expansion of the dimensionless strain energy $W$ at order 2  in terms of $I_1-3$ and $I_2-3$. Here, we consider the most general form for this expansion, without any restriction to a specific kind of constitutive equation: \begin{equation}
	W=\frac{1}{2}\beta(I_1-3)+\frac{1}{2}\left(1-\beta\right)(I_2-3)+\gamma_{11}(I_1-3)^2+\gamma_{12}(I_1-3)(I_2-3)+\gamma_{22}(I_2-3)^2\cdots,
\label{eqn : 3rd order W}
\end{equation}
where $\beta$, $\gamma_{11}$, $\gamma_{12}$ and $\gamma_{22}$ are constant parameters that depend on the material properties. For instance, $\beta=1$ and $\gamma_{11}=\gamma_{12}=\gamma_{22}=0$ for an incompressible neo-Hookean solid~\citep{Ogden1984,Macosko94},  and $\gamma_{11}=\gamma_{12}=\gamma_{22}=0$ for an incompressible Mooney-Rivlin solid \citep{Mooney1940,Rivlin1948}.

The unknown functions in Eq.~\ref{eqn : gu}-\ref{eqn : gq} are found by solving at order $\varepsilon^2$ the differential equations Eqs.~\ref{eqn : 3d CP0 brute}-\ref{eqn : 3d CP3 brute} with the boundary conditions Eqs.~\ref{eqn : 3d BC3 brute}.
Inserting Eqs.~\ref{eqn : definition of alpha2}-\ref{eqn : definition of xi} into the Cauchy-Poisson Eqs.~\ref{eqn : 3d CP0 brute}-\ref{eqn : 3d CP3 brute} and in the boundary condition Eqs.~\ref{eqn : 3d BC3 brute} at order $\varepsilon^2$ yields: 
\begin{eqnarray}
  g_{u1}(r)&=&\frac{\xi^2}{r_0}\left\{-\frac{r}{4r_0}(r_0 k)^2+\left(4 \frac{r}{r}-\frac{r^3}{r^3_0}\right) \frac{(kr_0)^4}{32}+\left(7 \frac{r}{r_0}-6 \frac{r^3}{r_0^3}+2 \frac{r^5}{r^5_0}\right) \frac{(kr_0)^6}{128}   +\cdots \right\} \nonumber \\
  g_{u3}(r)&=&\frac{\xi^2}{r_0}\left\{-\frac{r}{8r_0}( kr_0)^2-\left((2 \beta-9) \frac{r}{r_0}+3 \frac{r^3}{r^3_0}\right) \frac{(kr_0)^4}{32}+\left((116 \beta-230) \frac{r}{r_0} +54 \frac{r^3}{r_0^3}+(12 \beta-33) \frac{r^5}{r^5_0}\right) \frac{(kr_0)^6}{1152}   +\cdots \right\}\nonumber\\
  g_{u5}(r)&=&\frac{\xi^2}{r_0}\left\{\frac{r}{8r_0} (kr_0)^2-\frac{r^3}{16r_0^3}(kr_0)^4+\left((150 \beta-269) \frac{r}{r_0}+(20 \beta+198) \frac{r^3}{r_0^3}+(-30 \beta-19) \frac{r^5}{r^5_0}\right) \frac{(kr_0)^6}{1536}   +\cdots \right\}\nonumber\\
  g_{u9}(r)&=&\frac{\xi^2}{r_0}\left\{\frac{r}{8r_0} (kr_0)^2-\left(6 \frac{r}{r_0}-3 \frac{r^3}{r_0^3}\right) \frac{(kr_0)^4}{16}+\left((2 \beta+561) \frac{r}{r_0}+(60 \beta-422) \frac{r^3}{r_0^3}+(-26 \beta+111) \frac{r^5}{r^5_0}\right)\frac{(kr_0)^6}{1536}   +\cdots \right\}\nonumber
\end{eqnarray}

\begin{eqnarray}
  g_{v5}(r)&=&\frac{\xi^2}{r_0}\left\{-(\frac{r}{8r_0} (kr_0)^2-\left((150 \beta-269) \frac{r}{r_0}+(40 \beta-36) \frac{r^3}{r_0^3}+(-90 \beta+75) \frac{r^5}{r^5_0}\right) \frac{(kr_0)^6}{1536}   +\cdots \right\}\nonumber\\
  g_{v9}(r)&=&\frac{\xi^2}{r_0}\left\{-\frac{r}{8r_0} (kr_0)^2+\left(3 \frac{r}{r_0}-\frac{r^3}{r^3_0}\right) \frac{(kr_0)^4}{8}-\left((2 \beta+561) \frac{r}{r_0}+(-8 \beta-348) \frac{r^3}{r_0^3}+(-14 \beta+49)\frac{r^5}{r_0^5}\right) \frac{(kr_0)^6}{1536}+\cdots \right\}\nonumber
\end{eqnarray}

\begin{eqnarray}
  g_{w3}(r)&=&\frac{\xi^2}{r_0}\left\{-\frac{kr_0}{8}+(2 \beta-5)\frac{(kr_0)^3}{32}-\left((116 \beta-293)+(36 \beta-63) \frac{r^4}{r^4_0}\right) \frac{(kr_0)^5}{1152}   +\cdots \right\}\nonumber\\
  g_{w9}(r)&=&\frac{\xi^2}{r_0}\left\{-\frac{r^2}{8r_0^2} (kr_0)^3-\left((16 \beta-44) \frac{r^2}{r_0^2}+(-8 \beta+31) \frac{r^4}{r^4_0}\right) \frac{(kr_0)^5}{192}   +\cdots \right\}\nonumber
\end{eqnarray}

\begin{eqnarray}
  g_{q1}(r)&=&\xi^2k^2\left\{\frac{1}{4}-\left(3-8(\beta-1 -6 \gamma) \frac{r^2}{r_0^2}\right) \frac{(kr_0)^2}{16}\right. \nonumber\\ &&\left. -  \left((-12 \beta+29+256\gamma)+(64 \beta-120-832\gamma) \frac{r^2}{r_0^2}+(-12 \beta+33+320\gamma) \frac{r^4}{r_0^4}\right)\frac{(kr_0)^4}{128}   +\cdots \right\}\nonumber\\
  g_{q3}(r)&=&\xi^2k^2\left\{\left(\beta+1+4(6 \gamma-\beta+1) \frac{r^2}{r_0^2}\right) \frac{(kr_0)^2}{8}\right. \nonumber \\ && \left.-\left((-44 \beta+515+2304\gamma)+(-720 \beta+252+2880\gamma)\frac{r^2}{r_0^2}+(180 \beta-153-576\gamma) \frac{r^4}{r_0^4}\right) \frac{(kr_0)^4}{1152}   +\cdots \right\}\nonumber\\
  g_{q5}(r)&=&\xi^2 k^2\left\{(24 \gamma-4 \beta+7) \frac{r^2}{r_0^2} \frac{(kr_0)^2}{8}-\left((-258 \beta+489+2496\gamma) \frac{r^2}{r_0^2}+(-24 \beta-74-960\gamma) \frac{r^4}{r^4_0}\right) \frac{(kr_0)^4}{384}   +\cdots \right\}\nonumber\\
  g_{q9}(r)&=&\xi^2 k^2\left\{-(24 \gamma-4 \beta+7) \frac{r^2}{r_0^2}\frac{(kr_0)^2}{8}+\left((-218 \beta+289+960\gamma) \frac{r^2}{r_0^2}+(72 \beta-106-192\gamma) \frac{r^4}{r_0^4}\right)\frac{(kr_0)^4}{384}   +\cdots \right\}\nonumber
\end{eqnarray}
with $\gamma=\gamma_{11}+\gamma_{12}+\gamma_{22}$. The other functions $g$ defined in Eqs.~\ref{eqn : gu}-\ref{eqn : gq} are equal to zero.
Note that the boundary conditions Eqs.~\ref{eqn :  boundary ends 1}  at $Z=0$ and $Z=L$ are fulfilled at order $\varepsilon^2$.

\subsection{Amplitude equation} \label{sec : amplitude}
The Koiter method proceeds by inserting the expansion in Eqs.~\ref{eqn
: definition of alpha2}--\ref{eqn : definition of xi} into the
non-linear equilibrium written earlier in Eq.~\ref{eqn : general1} as
\begin{equation}
 \forall  \delta{\mathbf{t}},\quad D{\cal E}\left(\alpha_{\mathrm{c}} +\alpha_2 \varepsilon^2, 
 \mathbf{t}_{0}(\alpha) + \varepsilon\mathbf{t}_{1} + \varepsilon^2\mathbf{t}_{2} +\varepsilon^3\mathbf{t}_{3}+ \cdots\right)[\delta \mathbf{t}]=0,
\label{eqn : equilibrium condition total}
\end{equation}
where $\delta \mathbf{t}(r,\theta,z)$ is the set of virtual functions
$\bigl(\delta{u},\delta{v},\delta{w},\delta{q}\bigl)$ that represent
infinitesimal increments of the displacements (including the Lagrange
multiplier) satisfying the kinematic boundary conditions.  
Eq.~\ref{eqn : equilibrium condition total} is then expanded
order by order in $\varepsilon$~\cite{Heijden2009koiter,Triantafyllidis-Stability-of-solids:-from-2011,Chakrabarti2018}. Order $\varepsilon$ of Eq.~\ref{eqn : equilibrium condition total} yields the linear bifurcation problem :
\begin{equation}
  \forall \delta{\mathbf{t}},\quad   D^2{\cal E}\left(\alpha_{\mathrm{c}}\right)
  \left[\mathbf{t}_{1},\delta \mathbf{t}\right]=0.
  \label{eq:KoiterOrder2}
\end{equation}
Order $\varepsilon^{2}$ yields the  equations for the second-order correction $\mathbf{t}_{2}$. One obtains at $\varepsilon^{3}$ the equation:
	\begin{multline}
   \forall \delta{\mathbf{t}},\quad 
    D^2\mathcal{E}(\alpha_{\mathrm{c}},\mathbf{t}_{0}(\alpha_{\mathrm{c}}))\cdot 
    \left[\mathbf{t}_{3},\delta \mathbf{t}\right]+ 
    D^3\mathcal{E}(\alpha_{\mathrm{c}},\mathbf{t}_{0}(\alpha_{\mathrm{c}}))\cdot 
    \left[\mathbf{t}_{2},\mathbf{t}_{1},\delta \mathbf{t}\right]  \\
    {}+\alpha_2 \left.\frac{\mathrm{d} D^2\mathcal{E}(\alpha,\mathbf{t}_{0}(\alpha))}{\mathrm{d}\alpha}
    \right|_{\alpha=\alpha_{\mathrm{c}}}
    \cdot \left[\mathbf{t}_{1},\delta \mathbf{t}\right]
    +\frac{1}{6}
    D^4\mathcal{E}(\alpha_{\mathrm{c}},\mathbf{t}_{0}(\alpha_{\mathrm{c}}))\cdot 
    \left[\mathbf{t}_{1},\mathbf{t}_{1},\mathbf{t}_{1},\delta \mathbf{t}\right]
    = 0
    \textrm{.} 
	\nonumber
\end{multline}
    Upon insertion of the particular virtual displacement 
	$\delta \mathbf{t} = \mathbf{t}_{1}$, the first term cancels out 
	by Eq.~\ref{eq:KoiterOrder2} and we are left with
	\begin{equation}
 D^3\mathcal{E}(\alpha_{\mathrm{c}},\mathbf{t}_{0}(\alpha_{\mathrm{c}}))\cdot 
    \left[\mathbf{t}_{2},\mathbf{t}_{1},\mathbf{t}_1\right]
    +\alpha_2\left.\frac{\mathrm{d} D^2\mathcal{E}(\alpha,\mathbf{t}_{0}(\alpha))}{\mathrm{d}\alpha}
    \right|_{\alpha=\alpha_{\mathrm{c}}}
    \cdot \left[\mathbf{t}_{1},\mathbf{t}_1\right]
    +\frac{1}{6}
    D^4\mathcal{E}(\alpha_{\mathrm{c}},\mathbf{t}_{0}(\alpha_{\mathrm{c}}))\cdot 
    \left[\mathbf{t}_{1},\mathbf{t}_{1},\mathbf{t}_{1},\mathbf{t}_1\right]
    = 0
    \textrm{.} 
    \label{eqn : gateau4}
        \end{equation}
  $D^2{\cal E}\left[\delta
\mathbf{t};\mathbf{t}_1\right]$ denotes the second G\^ateaux
derivative of ${\cal E}$, which is a bi-linear symmetric form on the
increment $\mathbf{t}_1$ and on the virtual increment $\delta
\mathbf{t}$.  Similarly, $D^3{\cal E}\left[\delta
\mathbf{t};\mathbf{t}_1;\mathbf{t}_2 \right]$ is the third G\^ateaux
derivative (a tri-linear symmetric form).
 From Eqs.~\ref{eqn : definition of alpha2}--\ref{eqn : definition of xi}, the value of $\alpha_{2}$
finally allows to express the buckling amplitude $\varepsilon \xi$ as a function of 
the increment of the load $\alpha-\alpha_{c} = \alpha_{2}\varepsilon^{2}$.
The quantities appearing Eq.~\ref{eqn : gateau4} are calculated with the help of a symbolic calculation language from the explicit expression of the functions $f$ (linear order) and the functions $g$ (second order):
\begin{eqnarray}
  D^3\mathcal{E}(\alpha_{\mathrm{c}},\mathbf{t}_{0}(\alpha_{\mathrm{c}}))\cdot [\mathbf{t}_{2},\mathbf{t}_{1},\mathbf{t}_1]  &=&\frac{\mu \pi^2 r_0^2}{k} \left(\frac{\xi}{r_0} \right)^4\left\{ -\frac{3(kr_0)^4}{8}(16 \gamma-2 \beta+1 )+(kr_0)^6(2 \gamma-\frac{5}{16})+\cdots \right\},\\
  D^4\mathcal{E}(\alpha_{\mathrm{c}},\mathbf{t}_{0}(\alpha_{\mathrm{c}}))\cdot [\mathbf{t}_{1},\mathbf{t}_{1},\mathbf{t}_{1},\mathbf{t}_1  &=&\frac{\mu \pi^2 r_0^2}{k} \left(\frac{\xi}{r_0} \right)^4\left\{\frac{9(kr_0)^4}{2}(8 \gamma- \beta+1)-\frac{3 (kr_0)^6}{8}(32\gamma-6  \beta+7)    +\cdots \right\},\\
  \left.  \frac{\mathrm{d} D^2\mathcal{E}(\alpha,\mathbf{t}_{0}(\alpha))}{\mathrm{d}\alpha}\right|_{\alpha=\alpha_{\mathrm{c}}}&=&-\frac{\mu \pi^2 r_0^2}{k}\frac{1}{2} \left(\frac{\xi}{r_0} \right)^2. 
\end{eqnarray}
Eq.~\ref{eqn : gateau4} then yields the sought relation between
the scaled load increment $\alpha_2$ and amplitude $\xi$:
\begin{equation}
  \alpha_2=\kappa \frac{\xi^2}{r_0^2},
  \label{eqn : kappa2}
\end{equation}
with
\begin{equation}
  \kappa=\frac{3}{4} (kr_0)^4+\frac{3}{4}(\beta-2) (kr_0)^6+\left(\frac{27}{8} \gamma-\frac{3}{32} \beta^2-\frac{61}{96} \beta+\frac{105}{128}\right) (kr_0)^8+\cdots
  \label{eqn : kappa}
\end{equation}
Multiplying both sides of Eq.~\ref{eqn : kappa2} by $\varepsilon^{2}$ and identifying ({\it i}) the load
increment $\alpha-\alpha_{c}$ from Eq.~\ref{eqn : definition of alpha2}
and ({\it ii}) the true buckling amplitude $\zeta = \varepsilon\,\xi$, we find
the amplitude equation as
\begin{equation}
  \left( \frac{\zeta}{r_0}\right)^2=\frac{1}{\kappa}\left( \alpha-\alpha_c\right).
  \label{eqn : zeta order 2}
\end{equation}
\begin{figure}[!h]
\begin{center}
\includegraphics[width=0.55\textwidth]{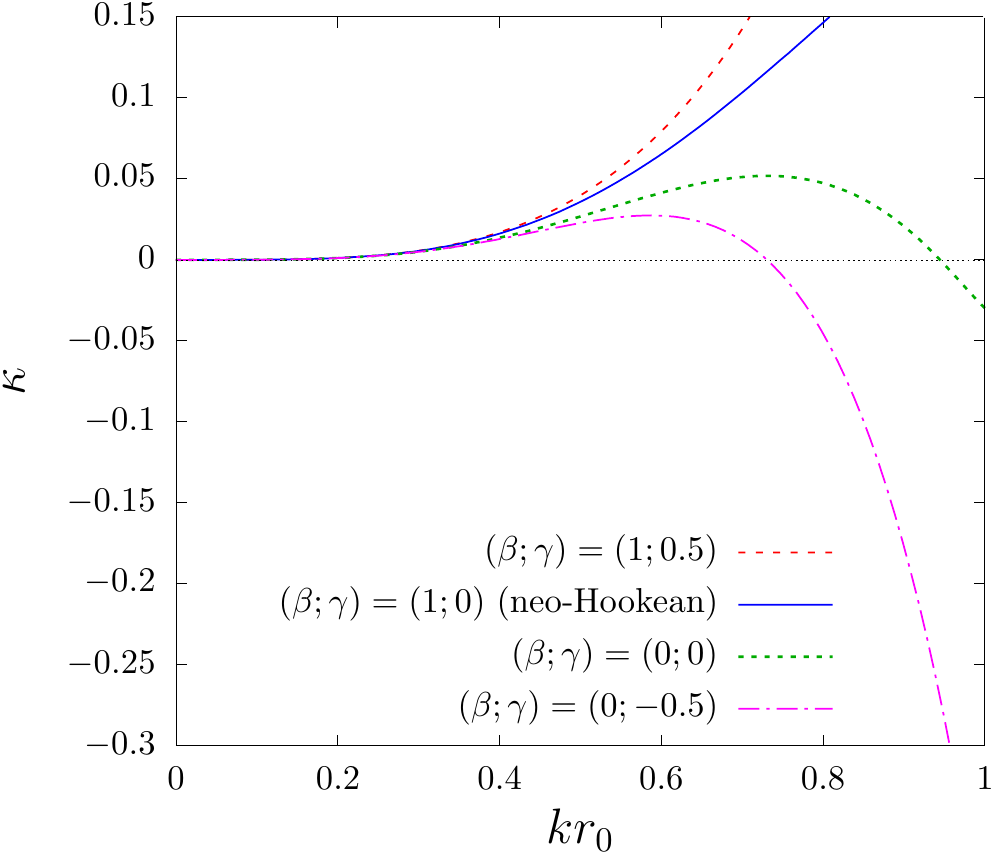}
\end{center}
\caption{$\kappa$ defined by Eq.~\ref{eqn : kappa} calculated from series expansions at order 30 in $kr_0$ for a neo-Hookean constitutive law $(\beta;\gamma)=(1;0)$; two Mooney-Rivlin constitutive laws  $(\beta;\gamma)=(0.5;0)$ and  $(\beta;\gamma)=(1;0)$ ; and a constitutive law with $(\beta;\gamma)=(0;-0.5)$.} \label{fig : kappa}
\end{figure}
$\kappa$ is plotted as a function of $kr_0$ for different constitutive laws in Figure \ref{fig : kappa}. The order in the expansion in $kr_0$ in Eq.~\ref{eqn : kappa}  is high enough to have no visible effect of it in this plot. Interestingly, $\kappa$ can be positive or negative, depending on the constitutive law and on the values of $kr_0$.  
Based on the amplitude equation in Eq.~\ref{eqn : zeta order 2}, the
bifurcation is super critical (continuous) if $\kappa>0$  ($\zeta=r_0\sqrt{(\alpha-\alpha_c)/\kappa}$) and sub-critical (discontinuous) otherwise ($\zeta=r_0\sqrt{(\alpha_c-\alpha)/\kappa}$).
The bifurcated branch is found above the critical load $\alpha_c$ in the super critical case, and below $\alpha_c$ in the sub critical case.  
\begin{figure}[!h]
\begin{center}
\includegraphics[width=0.55\textwidth]{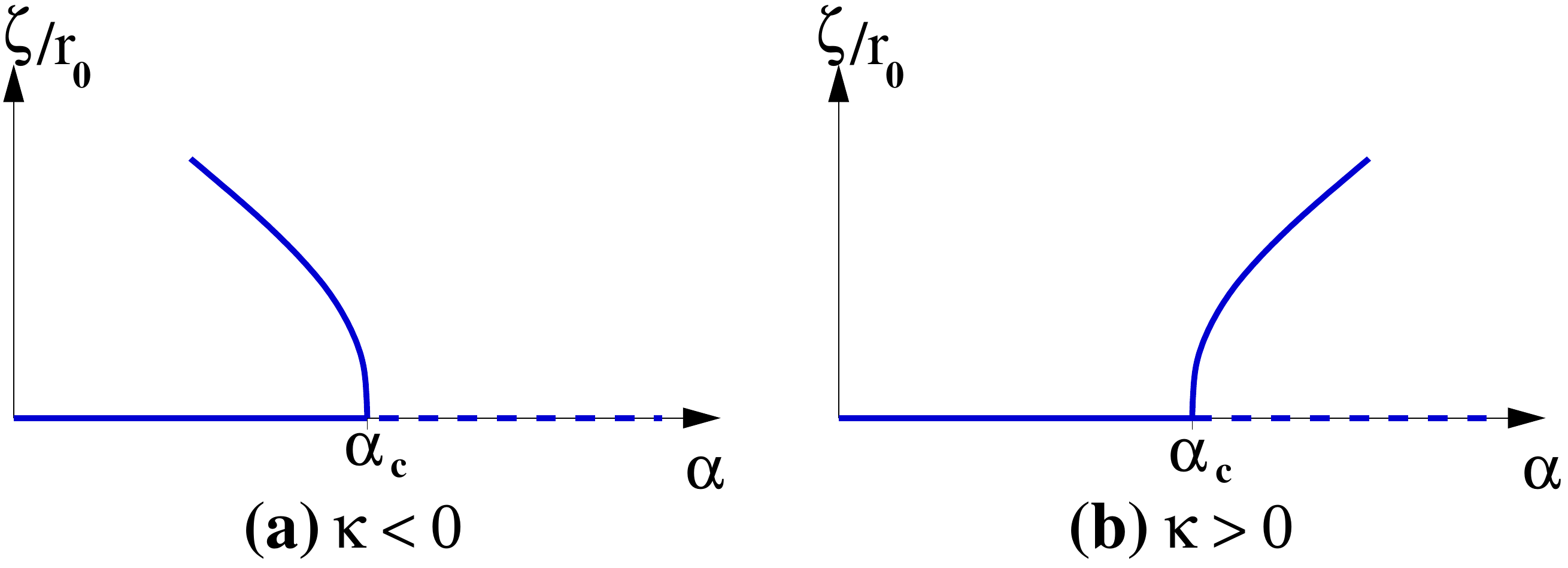}
\end{center}
\caption{Sketches of the buckling amplitude close to the bifurcation point. (a) for $\kappa<0$ the bifurcation is sub-critical. (b) for $\kappa>0$ the bifurcation is super-critical.} \label{fig : bifurcations}
\end{figure}

\section{Comparison with finite element simulations} \label{sec : FEM}

In this section, the complete non-linear problem defined by Eq.~\ref{eqn : general1} is implemented by using the open source tool for solving partial differential equations FEniCS \cite{Fenics2012}. The goal is to check whether the numerical simulations well capture the results of both the linear and the non-linear analysis of sections \ref{sec : linear}-\ref{sec : non linear}.

We consider a semicircular solid cylinder $\Omega$ of radius $r_0$ and height $2 \pi /k$. A Cartesian coordinates system ($x,y,z$) with the base vectors ($\mathbf{e}_x,\mathbf{e}_y,\mathbf{e}_z$) is chosen such that $(x,y,z) \in \Omega \Leftrightarrow \sqrt{x^2+y^2}\leq r_0$, $x \geq 0$ and $0 \leq z \leq 2 \pi / k$. An incompressible and isotropic elastic solid (mass density $\rho$ ; shear modulus $\mu$) occupying the domain $\Omega$ in its reference configuration is subjected to the action of the centrifugal volume force $\rho \omega^2 (x\mathbf{e}_x+y\mathbf{e}_y)$. 
The lateral surface ($\sqrt{x^2+y^2}= r_0$) of the cylinder is traction free, the displacements in the direction of $\mathbf{e}_x$ are set to zero for $x=0$ (so that $x=0$ is a plane of symmetry), and periodic boundary conditions along axis $z$ with wave number $k$ are implemented.\\

The displacement vector $\mathbf{u}$ is discretized using Lagrange finite elements with a quadratic interpolation, and the Lagrange multiplier $q$  with a linear interpolation. The non-linear problem in $\mathbf{u}-q$ is solved using a Newton algorithm based on a direct parallel solver (\texttt{MUMPS}). Quasi-static simulations are performed by setting $\mu=1$, $r_0=1$, and slowly varying $\alpha\equiv \rho \omega^2$ up to the desired value. For each $\alpha$ the displacement field and the Lagrange multiplier are computed. Simulations are carried out for different values of the wave number and different elastic constitutive laws.

\begin{figure}[!h]
\begin{center}
\includegraphics[width=0.45\textwidth]{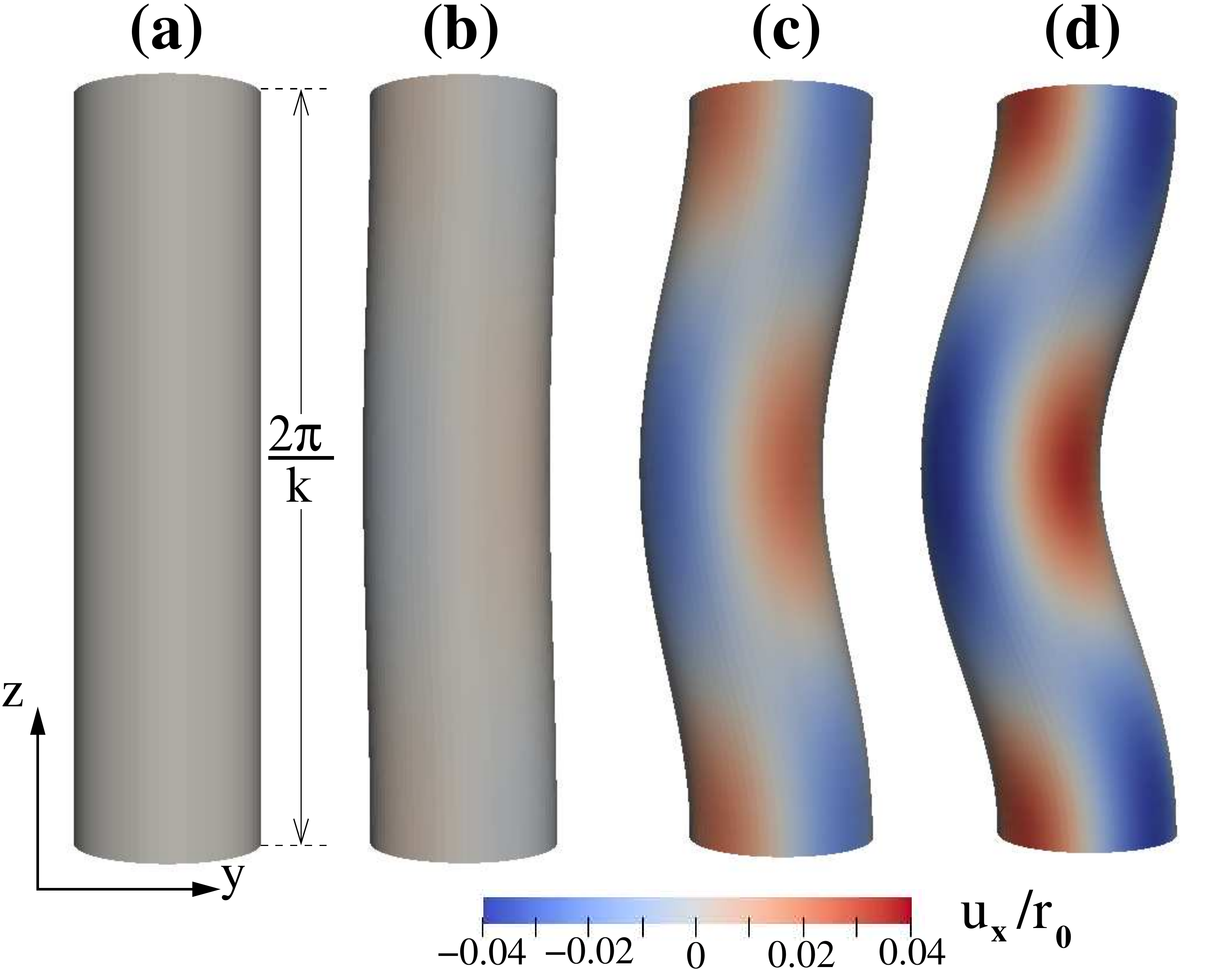}
\end{center}
\caption{Snapshots from FEM simulations of a neo-Hookean spinning cylinder with $kr_0=\frac{\pi}{4}$, for $\frac{\alpha-\alpha_c}{\alpha_c}=0$ (a), $\num{8.4e-4}$ (b), $\num{4.2e-2}$ (c) and $\num{0.1}$ (d). The buckling amplitudes are respectively equal to $0$ (a), $0.037$ (b), $0.29$ (c) and $0.45$ (d). Colors indicate the normalized displacement along $x$-direction, $u_x/r_0$.} \label{fig : snap}
\end{figure}

\begin{figure}[!h]
\begin{center}
\includegraphics[width=0.45\textwidth]{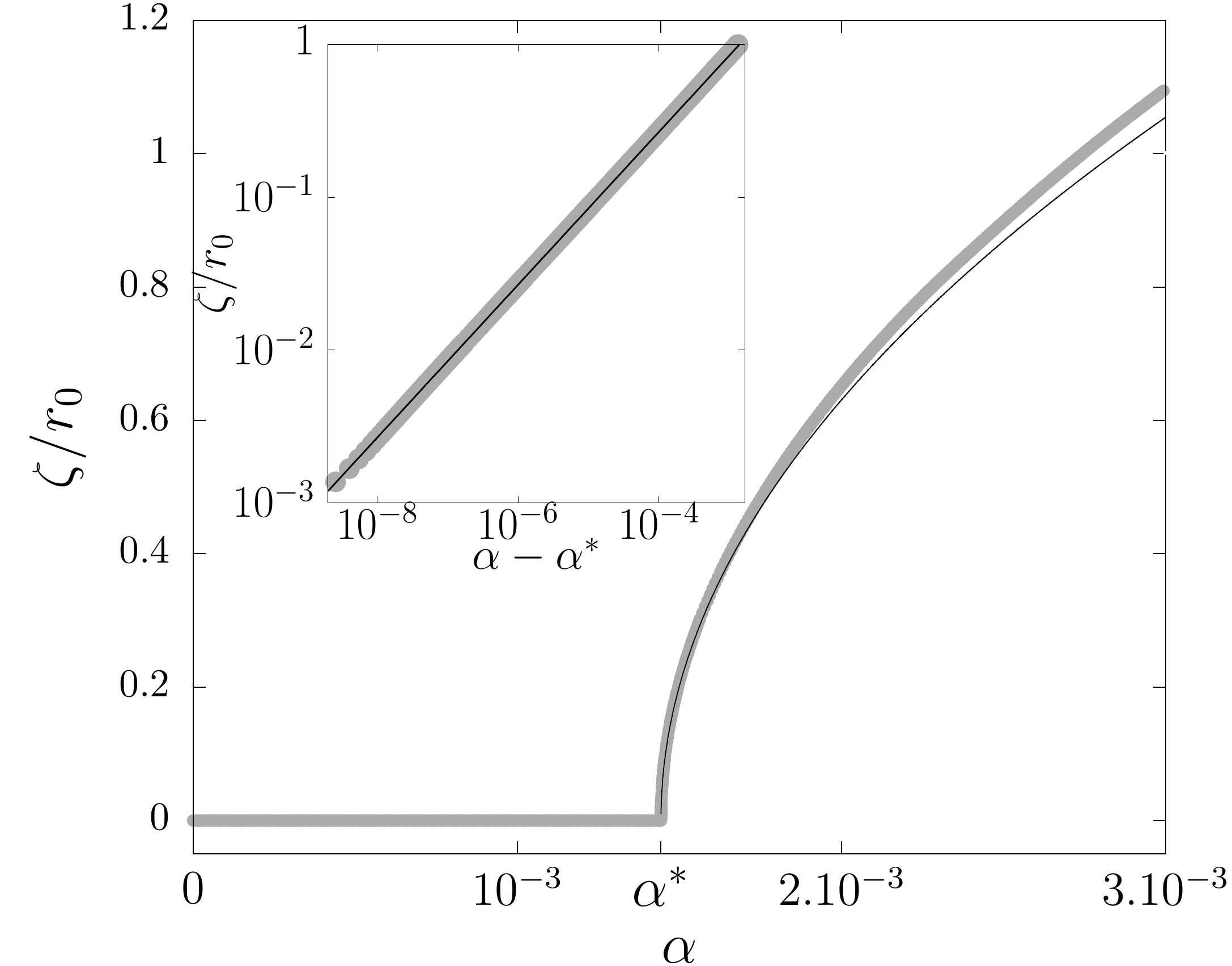}
\end{center}
\caption{Gray symbols: Normalized buckling amplitude $\zeta/r_0$ as a function of $\alpha$ for the neo-Hookean constitutive law, obtained by FEM simulations for $kr_0=\pi/15$. Solid line is obtained by fitting  $\zeta/r_0=\sqrt{(\alpha-\alpha^*)/\kappa^*}$ via $\alpha^*$ and $\kappa^*$: $\alpha^*=0.00144$ and $\kappa^*=0.00137$. Inset : same data plotted in log-scales.} \label{fig : p030}
\end{figure}

Starting from the undisturbed base system ($\mathbf{u}=0$), $\alpha$ is gradually increased with increments $\delta \alpha=1/100000$.  The deformation is almost zero until a critical value of $\alpha$ for which the deformation begins to increase (as a function of $\alpha$) abruptly. Due to the boundary conditions imposed in the simulations, these deformations are consistent with those investigated in Secs. \ref{sec : linear}-\ref{sec : non linear} (see Figure \ref{fig : snap}). Accordingly with the definition in Section \ref{sec : amplitude} of buckling amplitude $\zeta$,  the buckling amplitude in the simulations is computed as the maximum displacement of the material points located at the lateral boundary of the cylinder. The normalized buckling amplitude $\zeta/r_0$ computed from the simulations for a neo-Hookean constitutive law and the wave number $kr_0=\pi/15$ is plotted as a function of $\alpha$ in Figure \ref{fig : p030}. Fitting   $\zeta/r_0$ with the function $f(\alpha)$ defined by :
\begin{eqnarray}
  f(\alpha)=0 &\mbox{ for }& \alpha<\alpha^* \nonumber \\
  f(\alpha)=\sqrt{\left(\alpha-\alpha_c\right)/\kappa^*} &\mbox{ for }& \alpha>\alpha^*, \label{eqn : star}
\end{eqnarray}
for $\zeta/r_0<0.2$ gives values of the threshold $\alpha^*$ computed from the simulations, as well as the coefficient $\kappa^*$. 
\begin{figure}[!h]
\begin{center}
\includegraphics[width=0.5\textwidth]{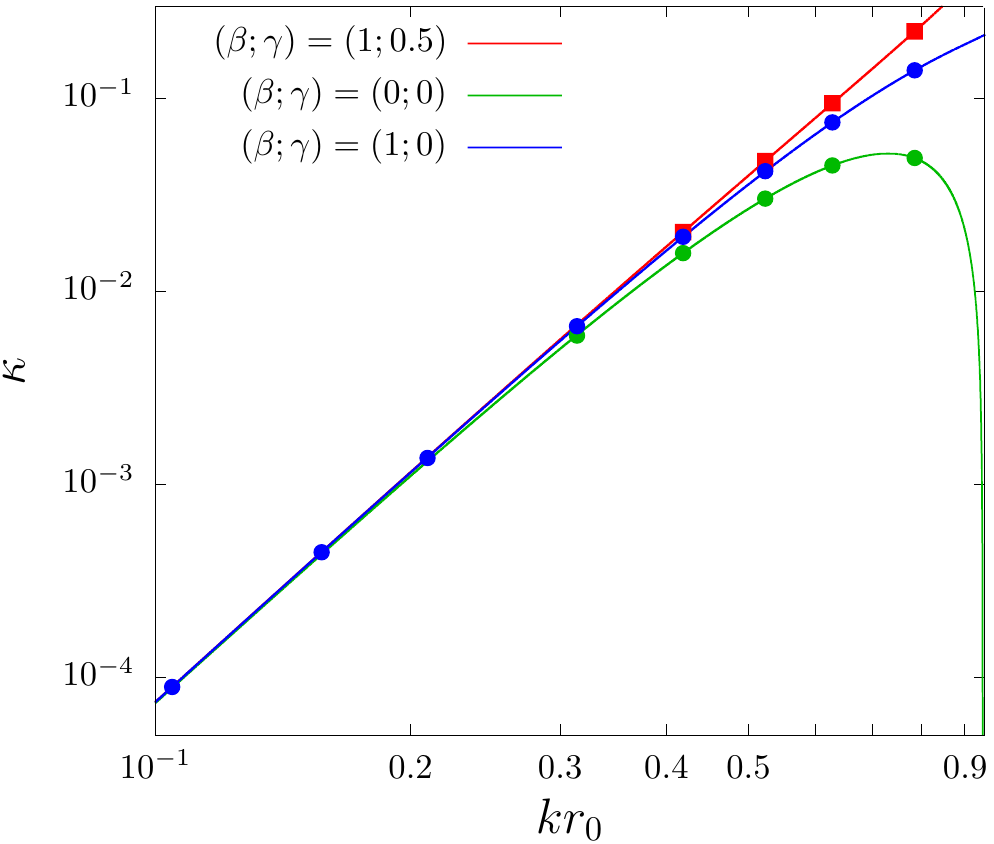}
\end{center}
\caption{$\kappa$ as a function of $kr_0$ in log-log scale for different constitutive laws. Solid lines result from of Eq.~\ref{eqn : kappa} (calculated at order 30 in $kr_0$; the plot would not change by increasing the order). Filled circles results from fits of the buckling amplitude calculated by FEM simulation, as presented in Figure \ref{fig : p030}. } \label{fig : kappa log}
\end{figure}

\begin{figure}[!h]
\begin{center}
\includegraphics[width=0.5\textwidth]{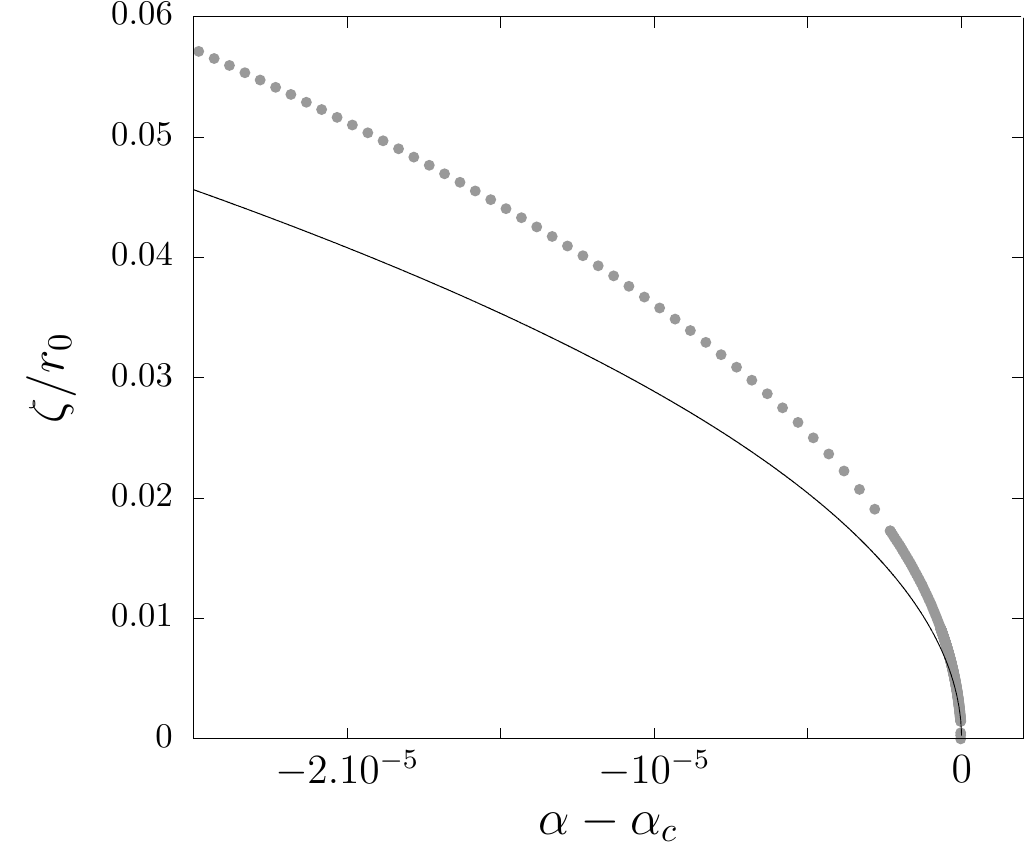}
\end{center}
\caption{Filled circles: Normalized buckling amplitude $\zeta/r_0$ as a function of $\alpha$ for a Mooney-Rivlin constitutive law with $\beta=0$, obtained by FEM simulations for $kr_0=\pi/3.25$.  Solid line is the prediction of the weakly non-linear theory of Section \ref{sec : non linear}.} \label{fig : q0065}
\end{figure}

In Figure \ref{fig : linear threshold}, $\alpha^*$ is plotted together with the theoretical prediction of the linear threshold, Eq.~\ref{eqn : alphac}. A comparison of $\kappa^*$ with the theoretical prediction based on the weakly non-linear analysis, Eq.~\ref{eqn : kappa}, is shown in Figure \ref{fig : kappa log} for different wave numbers and different constitutive laws. The good agreement between theory and simulations clearly validates the results of Sections \ref{sec : linear} and \ref{sec : non linear}.
In addition, the simulations show that the prediction of Eqs.~\ref{eqn : kappa}-\ref{eqn : zeta order 2} remains good for finite values of $\zeta/r_0$ (Figure \ref{fig : p030}). Discrepancies with the square root expression of the weakly non-linear analysis are barely observable in log-log scales (inset of Figure \ref{fig : p030}). For instance, for a neo-Hookean constitutive law and $kr_0=15\pi$, differences are smaller than 1\% for $\zeta/r_0<90\%$. \\
The sub-critical nature of the bifurcation, unveiled in Figure \ref{fig : kappa} for certain values of $\beta$ and $\gamma$ and certain values of the wave number, is also captured by the FEM simulations. In those cases, the load $\alpha$ has to be gradually decreased from a value larger than the instability threshold, and the buckling amplitude is found to grow as $\alpha$ continues to decrease below the critical load (Figure \ref{fig : q0065}). For these sub-critical bifurcations, the range of the load in which the buckling amplitude follows a square root law is more reduced compared to the super-critical case.  

\section{Concluding remarks} \label{sec : conclusion}

The non-linearities driving the buckling amplitude in the bifurcation of an initially straight spinning cylinder arise both from the geometry and the elastic response of the material. They simultaneously appear at order 2 in expansions with respect to the amplitude of the deformations.
The buckling amplitude has been calculated in the weakly non-linear regime for different wave numbers and for any  isotropic and isochoric constitutive law of the elastic material. Since the calculation relies on a Koiter expansion of the deformation calculated from a base undeformed configuration, the obtained analytic expression is limited to infinitesimal deformations. It has been complemented with numerical simulations, showing that the analytic expression is indeed relevant beyond the limit of the small infinitesimal deformations.

In the long wave length limit ($kr_0\ll 1$), $\alpha_c \sim \kappa \sim \frac{3}{4}(kr_0)^4$ (from Eqs.~\ref{eqn : alphac} and \ref{eqn : kappa}). Hence, $\zeta=r_0\sqrt{\alpha/\alpha_c-1}$. This formula differs from the expression proposed in \cite{Hosseini2013} and established in the long wave length limit through a one-dimensional model and by ignoring material non-linearities. This discrepancy shows that an approach based on Hookean elasticity for calculating the buckling amplitude is not relevant even in the long wave length limit. Indeed, calculating the buckling amplitude using reduction to one dimensional model would require a reduction consistent with the non-linear material constitutive law of the elastic material \cite{Audoly2019}.  

Non-linearities in the elastic material properties are a key ingredient for the study of the whirling instability: the buckling amplitude at the instability onset, and also the nature of the bifurcation (sub-critical or super-critical) depend on the coefficients appearing in the second order expansion of the strain energy density. Indeed, non-linear elastic properties are important in many systems or devices in which elastic bodies are subjected to finite deformations, as in soft robotics and surgery. As in the whirling instability investigated here, these deformations can be associated to instabilities \cite{Mora_softmatter2011,Mora_prl2014,Mora_softmatter2019} that lead to dramatic change in the system behaviour. The development of rigorous frameworks and methodologies for predicting, understanding and analysing these instabilities is then an important task.

The deformations considered in this paper being stationary, the equilibrium configurations have been analyzed by minimizing the total energy of the system, since energy dissipation processes are not relevant. A study of the issue of the transient regimes, {\it i.e.} the way the previously investigated steady states are reached, would required  more complex formulations in which the dissipative processes have to be accounted for together with the material and geometric non-linearities in dynamical equations. \\  

\noindent {\bf Acknowledgments:} Corrado Maurini is thanked for his help with FEniCS.\\
\vglue 0.05\textwidth
\noindent {\bf Conflict of Interest:} The author declares that he has no conflict of interest.


\end{document}